\documentstyle[prl,aps,twocolumn,epsfig]{revtex}
\begin{document}
\draft
\title
{What determines the spreading of a wave packet?}
\author{R. Ketzmerick$^{1,2}$,
K. Kruse$^{2,3}$,
S. Kraut$^{3}$, and
T. Geisel$^{1,2}$}

\address{
$^1$Institute for Theoretical Physics, 
University of California Santa Barbara, CA 93106, USA \\
$^2$Max-Planck-Institut f\"ur Str\"omungsforschung und 
Institut f\"ur Nichtlineare Dynamik der Universit\"at G\"ottingen,\\
Bunsenstra{\ss}e 10, D-37073 G\"ottingen, Germany$^*$ \\ 
$^3$Institut f\"ur Theoretische Physik und
SFB Nichtlineare Dynamik, \\
Universit\"at Frankfurt, D-60054 Frankfurt/Main, Germany\medskip\\
\parbox{14cm}{\rm
The multifractal dimensions $D_2^{\mu}$ and $D_2^{\psi}$
of the energy spectrum and eigenfunctions, resp., are shown to determine the
asymptotic scaling of the width of a spreading wave packet.
For systems where the shape of the wave packet is preserved the
$k$-th moment increases as $t^{k\beta}$ with $\beta=D_2^{\mu}/D_2^{\psi}$,
while in general $t^{k\beta}$ is an optimal lower bound.
Furthermore, we show that in $d$ dimensions asymptotically in time the
center of any wave packet decreases
spatially as a power law with exponent $D_2^{\psi}-d$ and present
numerical support for these results.
}}
\maketitle
\narrowtext

For the case of a free particle, the spreading of a quantum mechanical wave
packet is a textbook example. One finds that asymptotically the width
increases linearly in time and the corresponding energy spectrum is
absolutely continuous. In the case of a point spectrum, on
the other hand, there is asymptotically no increase of the width of a wave
packet, if the eigenfunctions are semiuniformly localized \cite{djls95}. 
Between these extremes flourishes the world of
quantum systems with 
fractal energy spectra and eigenfunctions. They include
systems studied in the
early days of quantum mechanics, like Bloch electrons in
a magnetic field \cite{peierls} as well as quasicrystals
\cite{shechtman} and disordered systems, e.g.\ at the
Anderson transition \cite{schreiber} or in the Quantum Hall regime
\cite{hs94}.
A natural question then arises: What determines their long-time
dynamical properties, e.g.\ the spreading of wave packets (quantum
diffusion) and the decay of temporal correlations?

Very few rigorous and general answers are known so far.
Temporal correlations decay as $t^{-D_2^{\mu}}$
\cite{kpg92},
where $D_2^{\mu}$ is the correlation dimension of the
spectral measure $\mu$ (i.e.\ the local density of states).
For the spreading of a wave packet, on the other hand,
two {\em inequalities} were derived by Guarneri.
The growth of the $k$-th moment $m_k(t) \sim t^{k\beta_k}$
is bounded from below by $D_1^{\mu} \leq \beta_k$
\cite{guarneri89+93}
and the entropic width grows faster than $t^{D_1^{\mu}}$
\cite{guarneri96},
where $D_1^{\mu}$ is another generalized dimension of the spectral measure.

In search for {\em equalities}, relating the growth of the moments
of a wave packet to fractal properties of the spectrum,
numerical studies of the Harper model
\cite{ha88a,geisel91},
the Fibonacci model
\cite{ha88,cop},
and the kicked Harper model
\cite{kickedharper}
were performed, suggesting, e.g.\ $\beta_2=D_0$, 
in agreement with a heuristic argument
\cite{guarneri89+93,geisel91},
where $D_0$ is the fractal (box-counting) dimension of the spectrum.
More recently, numerical studies
\cite{gm94,wilkinson}
showed that this simple relation does not hold exactly.
These studies also revealed that one often finds multiscaling in time
\cite{ek93}, 
i.e. $\beta_k$ varies with $k$. For a restricted class of systems, Mantica
showed that $\beta_k=D_{1-k}^\mu$, whereas in general it is at best
approximate \cite{mantica96}. The similar relation $\beta_k=D_{1-k}$,
now with the multifractal dimension of the ("global") density of
states, was proposed to hold after averaging the dynamics of {\em all} wave
packets started at different initial sites \cite{piechon96}. 
Therefore, the basic question, of what determines the spreading
of a {\em single} wave packet, still remains open.

In this paper we derive a partial answer to this question (Fig.~1), 
by including, for the first time, fractal properties of the eigenfunctions.
For systems where asymptotically the
$k$-th moment of a wave packet increases
proportional to $t^{k\beta}$, we show
\begin{equation} \label{beta}
\beta = D_2^{\mu}/D_2^{\psi} ,
\end{equation}
where $D_2^{\psi}$ is the correlation dimension of the
(suitably averaged) eigenfunctions.
For the more general case of multiscaling in time we derive an optimal
lower bound for positive moments under reasonable assumptions:
\begin{equation} \label{betak}
\beta_k \geq D_2^{\mu}/D_2^{\psi}.
\end{equation}
\begin{figure}[b]
\centering
\epsfig{figure=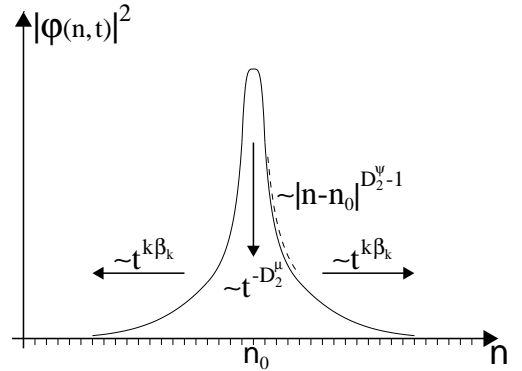,width=7.3cm}
\caption{\footnotesize
Quantum diffusion of a wave packet initially started at site $n_0$.
The staying probability decays as $t^{-D_2^{\mu}}$, whereas the $k$-th
moment increases as $t^{k\beta_k}$.
If the form of the wave packet remains unchanged (uniform scaling)
Eq.~(\protect\ref{beta}) leads to $\beta_k=\beta\equiv D_2^{\mu}/D_2^{\psi}$
for $k>0$.
For the case of multiscaling dynamics we find the lower bound
$\beta \leq \beta_k$ for $k>0$ (Eq.~(\protect\ref{betak})).
In any case, the center of the wave packet spatially decreases as
$|n-n_0|^{D_2^{\psi}-1}$ for $|n-n_0|\ll t^\beta$.}
\end{figure}
We verify these results numerically for the on-site Fibonacci model and
the Harper model, where it is
a much better lower bound than the previously obtained
$D_1^{\mu}$ \cite{guarneri89+93},
which did not make use of fractal properties of the
eigenfunctions. Surprisingly, we find from these dynamical properties
that in $d$ dimensions asymptotically in time the center of any wave
packet decreases spatially like a power law with an exponent
$D_2^{\psi}-d$ (see Fig.~1).

The moments of a wave packet $\varphi(n,t)$, initially located on site $n_0$,
are given by $m_k(t)=\sum_{n\neq n_0} |n-n_0|^k |\varphi(n,t)|^2$,
where for convenience we use a finite one-dimensional lattice.
To derive the scaling properties of the moments we need a relation between
the integrated wave packet
\begin{equation}
g(x,t) = \sum_{|n-n_0|<x}|\varphi(n,t)|^2
\end{equation}
and the spectral function \cite{chalker}
\begin{equation}\label{S}
S(x,\omega) = \sum_{\stackrel{i, j}{|E_i-E_j|<\omega}}\sum_{|n-n_0|<x}
	       \psi_i^*(n_0)\psi_j(n_0)\psi_i(n)\psi_j^*(n)
\end{equation}
with $E_i$ and $\psi_i$ denoting eigenenergies
and eigenfunctions, respectively. 
For systems described by a real symmetric matrix these are simply related by
\begin{equation} \label{ggleichS}
g(x,t) = \int_0^\infty d\omega \cos\omega t \frac{d}{d\omega} S(x,\omega)
\end{equation}
and
\begin{equation}\label{Sgleichg}
S(x,\omega) = \frac{2}{\pi}\int_0^\infty dt \frac{\sin\omega t}{t} g(x,t) .
\end{equation}

For the special case $x=1$ 
the spectral function simplifies and we assume the
following scaling behavior \cite{remark1}:
\begin{equation}\label{Sx} 
S(x=1,\omega) =
\sum_{\stackrel{i, j}{|E_i-E_j|<\omega}} |\psi_i(n_0)|^2 |\psi_j(n_0)|^2
\sim \omega^{D_2^\mu}
\end{equation}
and
\begin{equation}\label{Som}
S(x,\omega = \omega_{\textrm{min}}) 
\sim x^{D_2^\psi},
\end{equation}
where $\omega_{\textrm{min}}$ is the smallest energy scale where the scaling of
Eq.~(\ref{Sx}) still holds for a given finite system. Physically, 
$\omega_{\textrm{min}}$ is the inverse of the time when a wave packet reaches 
the boundary and when the power-law moment growth is modified.
$D_2^\mu$ is the (standard) 
correlation dimension of the spectral measure and $D_2^\psi$ denotes the
correlation dimension of the averaged eigenfunctions \cite{D2}.

We will first study the special case that the overall shape of the wave packet
stays the same during its time evolution, with, of course, a scaling in
width and amplitude. Assuming a power-law scaling one therefore has 
a function $G$ of a single variable defined by
$G(x/t^{\beta}) = g(x,t)$.
This is equivalent to the property that
the positive ($k \geq 0$) moments scale as 
$m_k(t) \sim t^{k\beta}$, which we call uniform scaling. 
From Eq.~(\ref{Sgleichg}) and the properties of $G$ we
find that $S(x,\omega)$ is a
function of $x\omega^{\beta}$ only. Together with the scaling
behaviors of Eqs.~(\ref{Sx}) and (\ref{Som}) one easily finds
the result $\beta = D_2^{\mu} / D_2^{\psi}$ (Eq.~(\ref{beta})).
From Eq.~(\ref{ggleichS}) we can thus even determine 
all negative moments: 
$m_k(t) \sim t^{k\beta}$ for $k \ge -D_2^\psi$,
whereas
$m_k(t) \sim t^{-D_2^\mu}$ for $k \le -D_2^\psi$.

Secondly, we now want to study the case of multiscaling dynamics.
We cannot do this in the most general form, but need two reasonable
assumptions. We first assume that for $x\omega^{\beta} \ll 1$ the
spectral function $S(x,\omega)$ is determined by its limiting scaling
behaviors (Eqs.~(\ref{Sx}) and (\ref{Som})), namely
$S(x,\omega) \sim x^{D_2^{\psi}} \omega^{D_2^{\mu}}$ \cite{remark2}.
Using Eqs.~(\ref{ggleichS}) and (\ref{Sgleichg})
this implies uniform scaling for the
center of the wave packet only, i.e.\
$g(x,t)=G(x/t^{\beta})$ for $x \ll t^{\beta}$. In fact, it follows
$g(x,t) \sim x^{D_2^\psi}t^{-D_2^\mu}$ for $x \ll
t^\beta$. Remarkably, this even determines the shape of the center of
the wave packet, namely, 
an algebraic decay $|n-n_0|^{D_2^\psi-1}$ for $|n-n_0| \ll
t^\beta$. In
addition, we assume that outside this region it decays faster than
$|n-n_0|^{D_2^\psi-1}$, which it certainly has to do asymptotically in
order to stay normalized. Under these assumptions,
it follows that for a large enough time $T$ there exists an $X$
such that $g(x,T) \propto x^{D_2^{\psi}}$ for $x \leq X$ and
that for all $t > T$
with $g(\xi(t),t)=g(X,T)$ the relation $\xi(t)/t^{\beta} \geq X/T^{\beta}$
must hold.
Using a result by Guarneri and Mantica \cite{guarneri89+93},
stating that all $\beta_k$ (for $k \geq 0$) 
are larger than the scaling exponent
of $\xi(t)$, it then follows
that all positive moments scale as $t^{k\beta_k}$ with
$\beta_k \geq \beta$.
Therefore $\beta$ is a lower bound on the scaling of the positive
moments that incorporates multifractal properties of the eigenfunctions.
In fact, for negative moments ($k<0$) we find that $\beta \geq \beta_k$ 
is an upper bound. In that sense $\beta$ is an optimal lower bound for
all positive moments.

Furthermore, note that this analysis can be readily extended to systems in
higher dimensions $d$ by 
appropriately generalizing the definitions of $g$ and $S$.
Under the condition $D_2^{\mu} < 1$ Eqs.~(\ref{beta}) and (\ref{betak})
remain unchanged and the spatial decay of the center of the wave packet
has an exponent $D_2^{\psi}-d$. Finally, this formalism can easily
describe the averaged dynamics of wave packets started at different
initial sites $n_0$. To this end we introduce an average in
Eqs.~(\ref{Sx}) and (\ref{Som}) defining exponents
$\bar{D}_2^\mu$ and $\bar{D}_2^\psi$, which replace $D_2^\mu$
and $D_2^\psi$ in the above results \cite{remark3}.

One can try to understand the derived results in an intuitive way: The
temporal decay of the center of the wave packet is known to be given
by $t^{-D_2^\mu}$ \cite{kpg92}. Naively, normalization then requires a
spreading described by an exponent $\beta=D_2^\mu / d$ in $d$
dimensions. If the spreading, however, takes place in a space with an
effectively reduced dimension $D_2^\psi$ instead of $d$ \cite{hs94},
we have $\beta=D_2^\mu / D_2^\psi$. Similarly, the integrated wave
packet $g(x,t)$, which usually increases initially as $x^d$, here
increases with $x^{D_2^\psi}$ instead. This leads immediately to a
power-law decay $|n-n_0|^{D_2^\psi -d}$ for the center of the wave
packet in the $d$-dimensional embedding space.

In the remainder, we want to give examples and
numerical evidence supporting the above analysis \cite{remark3a}.

For disordered systems at the metal-insulator transition in
2 (with strong magnetic field or symplectic symmetry) and 3 dimensions
Eq.~(\ref{beta}) 
is already known.
These systems are simpler in the sense that the fractal dimension
$D_0$ of the spectrum is 1 and it was shown that $D_2^{\psi} = d
D_2^{\mu}$ holds 
\cite{hs94,schweitzer95,brandes96}, where $d=2,3$ is the spatial
dimension.
On the other hand, 
$\beta=1/d$ follows from the non-fractal value of $D_0$ using 
heuristic arguments, as in Refs.~\cite{guarneri89+93,geisel91,cop},
and was numerically confirmed for the second moment
\cite{hs94,ok97}. The prediction that the asymptotic
shape of the wave packet follows the power law $|n-n_0|^{D_2^\psi -d}$
waits to be observed in disordered systems.

We now consider the on-site Fibonacci chain, which is a
one-dimensional model of a quasicrystal \cite{shechtman}
given by a tight-binding Hamiltonian
where the on-site energy $V_n$ takes the values $+V$ and $-V$, 
arranged according to the Fibonacci sequence.
Figs.~2 and 3 show that the ratio
$\beta=D_2^{\mu}/D_2^{\psi}= 0.477$ coincides, within the
numerical accuracy, with the scaling behavior of the moments,
thus confirming Eq.~(\ref{beta}) \cite{uni}. 

\begin{figure}[b]
\centering
\epsfig{figure=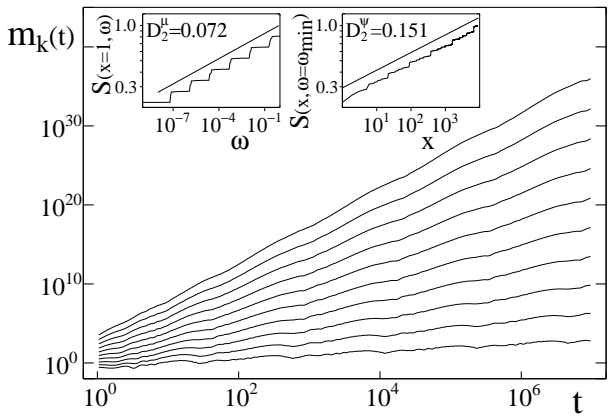,width=8.3cm}
\caption{\footnotesize
The moments $m_k(t)$ for $k=1, 2, \dots, 10$ of a wave packet initially
started at a symmetric site of a Fibonacci chain with $V=2$ and 17711
sites. The insets show the determination of $D_2^{\mu}=0.072$ (left) and
$D_2^{\psi}=0.151$ (right) from the limiting scaling behavior of
$S(x,\omega)$ yielding $\beta=D_2^{\mu}/D_2^{\psi}=0.477$,
which is much larger than the information dimension $D_1^{\mu}=0.153$.
}
\end{figure}

\begin{figure}[b]
\centering
\epsfig{figure=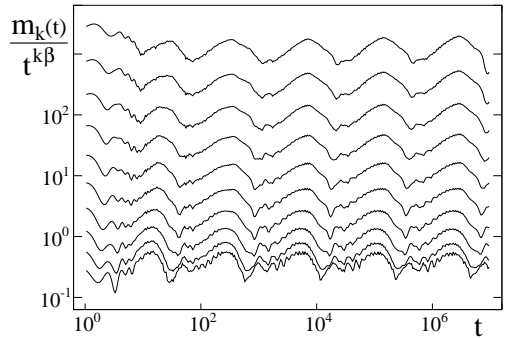,width=6.8cm}
\caption{\footnotesize
The scaled moments $m_k(t)/t^{k\beta}$ of a Fibonacci chain
corresponding to the parameters of Fig.~2
are (on average) constant, thus confirming Eq.~(\protect\ref{beta}).
}
\end{figure}

\begin{figure}[b]
\centering
\epsfig{figure=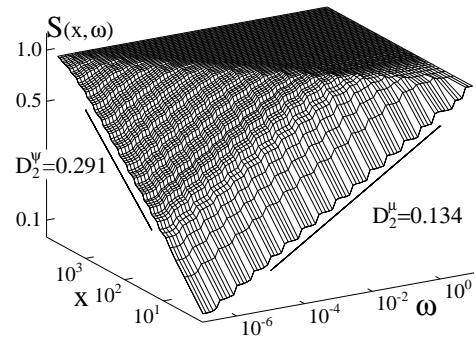,width=6.8cm}
\caption{\footnotesize
The spectral function $S(x,\omega)$ for the Harper model with
$\sigma=6765/10946$, an approximant of the golden mean, and $\nu=0$ of a
wave packet initially
started at $n=0$.
It shows the scaling behavior $\sim x^{D_2^{\psi}} \omega^{D_2^{\mu}}$
for small $x$ and $\omega$, as we have it assumed for deriving
Eq.~(\protect\ref{betak}).
(Deviations close to $S(x,\omega)=1$ might be the cause for multiscaling in
time.)
We find $D_2^{\mu}=0.134$ and $D_2^{\psi}=0.291$.
}
\end{figure}

\begin{figure}[b]
\centering
\epsfig{figure=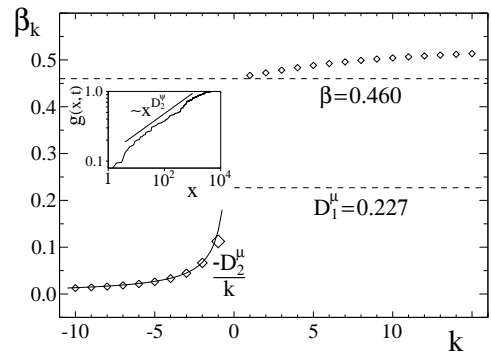,width=6.8cm}
\caption{\footnotesize
The values $\beta_k$ (diamonds) of the power-law scaling moments
$m_k(t) \sim t^{k\beta_k}$ of the Harper model
with $\sigma=10946/17711$ and $\nu=0$ of a
wave packet initially started at $n=0$.
One can see that $\beta=D_2^{\mu}/D_2^{\psi}=0.460$ is a much better
lower bound for all positive moments than $D_1^{\mu}=0.227$.
For negative moments, $k\leq -1$, we find $k\beta_k= -D_2^{\mu}$.
The inset shows the power-law
behavior of the integrated wave packet $g(x,t=3\times 10^6) \sim
x^{D_2^\psi}$ corresponding to a power-law decay $|\varphi(n,t)|^2\sim
|n-n_0|^{D_2^\psi -1}$.
}
\end{figure}

In general, though, quantum systems with a fractal energy spectrum generate
multiscaling dynamics \cite{gm94,wilkinson,ek93}. 
As an example we consider the Harper model
\cite{harper55}, describing an electron in a two-dimensional
periodic potential and a perpendicular magnetic field \cite{peierls}.
For $\sigma$ the number of magnetic flux quanta per
unit cell it is given by a tight-binding
Hamiltonian, where now $V_n = 2\cos ( 2\pi\sigma
n +\nu)$ holds. 
Fig.~4 shows that the scaling assumption for $S(x,\omega)$ is
fulfilled \cite{remark4} and Fig.~5 shows that $\beta \leq \beta_k$
is a good lower bound for the positive moments. We have verified
$\beta \le \beta_k$ also for wave packets started at other sites $n_0$
as well as for a few other irrational $\sigma$.
In fact, in all cases $\beta$ considerably improves the lower bound
$D_1^{\mu}$ deduced from spectral properties \cite{guarneri89+93}.
In addition, we have confirmed the prediction
for the shape of the center of the wave packet, namely a power-law
decay with exponent $D_2^{\psi}-1$, i.e. $g(x,t) \sim x^{D_2^\psi}$
(Fig.~5 inset).

This work was supported in part by the Deutsche Forschungsgemeinschaft
and by the NSF under Grant No.~PHY94-07194. 
We enjoyed many discussions with I.~Guarneri, Y.~Last, and G.~Mantica.


\begin{references}
\bibitem{remark1}
From Eq.~(\ref{Sx}) follows for the temporal correlation decay $C(t)
\equiv \frac{1}{t}\int_0^t dt^\prime \; g(1, t^\prime ) \sim
t^{-D_2^\mu} $ as was shown in Ref.~\cite{kpg92}.
\bibitem{remark2}
Under this condition the value of $D_2^{\psi}$ in Eq.~(\ref{Som}) does
not depend on the exact choice of $\omega_{\textrm{min}}$, as long as it is
in the scaling regime of $S(x,\omega )$.
\bibitem{remark3}
For the Harper model with $\sigma$ the golden mean we find
$\bar{D}_2^\mu = 0.265$ explaining the averaged temporal correlation
decay $\bar{C}(t)\sim t^{-0.25}$ found in Ref.~\cite{ek93}.
\bibitem{remark3a}
Ref.~\cite{djls95} gives a surprising example showing that a point spectrum
with exponentially localized eigenfunctions can coexist with moment growth 
at rates that are arbitrarily close to ballistic. Unfortunately, we cannot 
numerically check our results for this model due to the specially chosen
irrational.
\bibitem{uni}
If the on-site Fibonacci model for the chosen initial condition
were not a uniform scaling system, but a very weakly multiscaling system,
the data would still confirm the more general Eq.~(\protect\ref{betak})
within the numerical accuracy.
\end{references}
\end{document}